\documentclass[twocolumn,epsf,prl,amsmath,amssymb,showpacs]{revtex4}
\usepackage{graphicx}

\begin{document}
\title{Confined states  and direction
-dependent  transmission in graphene quantum wells}
\author{J. Milton Pereira Jr., V. Mlinar, F. M. Peeters}
\address{Department of Physics, University of Antwerp, Groenenborgerlaan 171, B-2020 Antwerpen}
\author{P. Vasilopoulos}
\address{Department of Physics, Concordia University, Montreal, Quebec, Canada H3G 1M8}

\begin{abstract}
We report the existence of confined massless fermion states in a
graphene quantum well (QW) by means of analytical and numerical
calculations. These states show an unusual quasi-linear dependence
on the momentum parallel to the QW: their number depends on the
wavevector and is constrained by electron-hole conversion in the
barrier regions. An essential difference with non-relativistic
electron states is a mixing between free and confined states at
the edges of the free-particle continua, demonstrated by the
direction-dependent resonant transmission across a potential well.
\end{abstract}
 \maketitle

%\vspace*{-0.7cm}
Recent studies have demonstrated the production of stable,
ultrapure, two-dimensional (2D) carbon crystals, also known as
graphene \cite{Novo3,Novo2,Zhang1}. These 2D crystals possess
unusual properties, such as unconventional quantum Hall effect
\cite{Zheng,Novoselov,Sharapov,Zhang} and a strong electric-field
effect \cite{Zhang2}. A large part of these new properties are a
consequence of the linear (in wavector) energy spectrum near the
Fermi energy and are expected to lead to a new class of carbon- or
graphene-based nanoelectronic devices. Previous theoretical
studies of relativistic fermions interacting with strong fields
have indicated that the quantum behavior of the particles may
differ considerably from the non-relativistic case
%\cite{High,Petrillo}
\cite{High}. In this paper we investigate the nature of electron
states in graphene  QWs and their quantized spectrum.

Graphene layers consist of a honeycomb lattice of covalent-bond
carbon atoms, which can be treated as two interpenetrating
triangular sublattices, %usually
labelled A and B, and are often discussed in terms of unrolled,
single-wall carbon nanotubes. The low-energy band structure of
graphene is gapless and the corresponding electronic states are
found near two cones located at unequivalent corners of the
Brillouin zone \cite{Wallace}. The low-energy carrier dynamics is
equivalent to that of a
 2D gas of  massless charged fermions.
Their behavior is governed by the 2D Dirac Hamiltonian
\cite{Semenoff,Kopele},
\begin{equation}
{\mathcal H}=v_F(\vec{\sigma}\cdot \hat{\mathbf p}),
\end{equation}
where the pseudospin matrix $\vec {\sigma}$ has components given
by Pauli's matrices, $\hat{\mathbf p} = (p_x,p_y)$ is the momentum
operator, and $v_F$  the effective
 speed of light of the system, which in this case corresponds to the
Fermi velocity $v_F \approx 1\times 10^6$ m/s. The Hamiltonian (1)
acts on the states represented by the two-component spinors $\Psi
= [\psi_A \, , \, \psi_B]^T$, where $\psi_A$ and $\psi_B$
represent the envelope functions associated with the probability
amplitudes at the respective sublattice sites of the honeycomb
graphene structure.  The low-energy spectrum of free carriers is
$E=\pm \hbar v_F(k_x^2+k_y^2)^{1/2}$, with $k_x$ and $ k_y$ the
wavevector along the $x$ and $y$ axes, in the vicinity of the
cones at the Brillouin zone; the $+$ ($-$) sign refers to
electron (hole) bands. %, respectively.
 Equation (1) also
implies that the carriers are chiral particles, with the
pseudospin aligned parallel (antiparallel) to the direction of
propagation of the electrons (holes).

Representing the effect of an external electrostatic field by
%can be obtained by means of
an external potential $U$  and including a diagonal effective
mass-like term $m\, v_F^2$ leads to the Dirac equation
\begin{equation}
[v_F(\vec {\sigma}\cdot \hat{\mathbf p})+m\, v_F^2 \sigma_z] \Psi
= (E-U)\Psi.
\end{equation}
The term $\propto m\, v_F^2$ creates a gap in the dispersion and
may arise from
spin-orbit interaction or %as consequence of
from the coupling between the graphene layer and the substrate
\cite{Kane}. For  a circularly symmetric potential with $m = 0$,
the solutions inside the potential well match free-particle
solutions outside, therefore ruling out bound states \cite{Vinc}.
This is caused by the conservation of the chirality in the
interaction with the potential and the absence of a gap in the
spectrum and can be understood as a manifestation of a
relativistic tunneling effect first discussed by Klein
\cite{Klein,Calogero} for one-dimensional (1D) potentials, in
which Dirac fermions can propagate to hole states across a steep
potential barrier without any damping. For massless particles this
tunnelling is expected to occur for any value of $U_0$. However,
as we show
below, for a 1D potential a finite %non-zero
value of the momentum parallel to
 the potential barrier can suppress this tunnelling and
thus allow the confinement of electrons.
%%%%%%%%%%%%%%%%%%%%%%%
Very recent studies have demonstrated the confinement of electrons
in a graphene strip \cite{Peres}. In this case, in order to obtain
the confinement the authors assumed a position-dependent effective
mass for the particles. This assumption does not permit the
observation of Klein tunnelling and of the momentum-dependent
reflection and transmission. Therefore, the confinement in this
case is qualitatively different from ours specified below. In
order to demonstrate the confinement in an electrostatic quantum
well, we  consider a zero or constant effective mass throughout
the system and first  a 1D square-well potential $U(x)=
U_0\,\theta(|x|-L/2)$, $U_0 > 0$, cf. Fig. 1, which allows an %direct
analytical solution for the eigenstates and sheds light on some
general features of the problem. Later on, we consider a parabolic
confinement.
%%%%%%%%%%%%%%%%%%%%%%%

With momentum conservation in %
%along
the $y$ %axis
direction,
we look for %assume
solutions in the form %of the type
$\psi_C(x,y)=\phi_C(x)e^{ik_y y}$,  C=A, B, %and
%$\psi_B(x,y)=\phi_B(x)e^{ik_y y}$,
and obtain
\begin{eqnarray}
%\frac{
d \phi_B/d \xi + \beta\phi_B &=& i(\epsilon-u-\Delta)\phi_A,\\
&{}&\cr %\frac{
d \phi_A/d \xi - \beta\phi_A &=& i(\epsilon-u+\Delta)\phi_B,
\end{eqnarray}
where $\xi = x/L$, $\beta = k_y L$, $\epsilon = %\frac{
E\,L/\hbar
v_F$, $u = %\frac{
U(x)\, L/\hbar v_F$ and $\Delta = m\, v_F
\,L/\hbar$ (for graphene %we have
$\hbar \, v_F = 0.539$ eV nm).
\begin{figure}
\centering{\resizebox*{!}{4.45cm}{\includegraphics{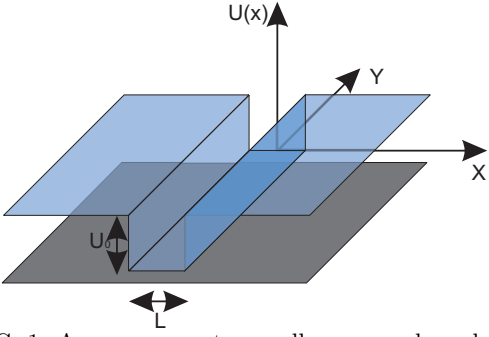}}}
\vspace{-0.5cm}
\caption{%Schematic depiction of a
A square quantum well on a graphene layer.} \label{fig:f1}
\end{figure}
Decoupling Eqs. % uations
(3) and (4) gives %can be readily decoupled. The result %, which gives,
 for $\phi_A$ the result
\begin{eqnarray}
&&\hspace*{-0.5cm} \frac{d^2 \phi_A}{d
\xi^2}+\frac{u'}{(\epsilon-u+\Delta)}\frac{d \phi_A}{d \xi}  \cr
&& -\Bigl[\beta^2 + \beta
\frac{u'}{(\epsilon-u+\Delta)}-(\epsilon-u)^2+\Delta^2
\Bigr]\phi_A=0,
\end{eqnarray}
where $u'$ is the derivative of the potential. For
%In the particularcase of
a square well, these derivatives are %found to be
Dirac delta functions.

The character of the solutions depends on the value of $\beta$,
which determines the sign of the last term
on the left %right-hand
side of Eq. (5). The solutions are of three types: (i) traveling
waves, which describe free electrons, free holes, as well as mixed
states that occur due to the Klein tunneling of electrons to holes
outside the potential well; (ii) standing waves, which for
 massless fermions arise only from
finite %non-zero
values of $\beta$ above an energy-dependent cut-off and decay
exponentially in the barrier regions; (iii) tunneling waves, which
are oscillatory outside the well whereas inside it they are
combinations of exponentials with real exponents; these correspond
to holes that undergo ordinary tunneling across the potential
well. Type (ii) solutions occur in energy and wavevector ranges
for which there are no hole states available at the barrier
regions. This suppresses the Klein tunneling, % mechanism,
since it depends on the electron-hole conversion at the interface.

In this work we focus on type (ii) solutions which describe
electron
states confined across %side
the well and propagating along it. % the QW.
Their energies are in the region delimited by the curves $E =
[(\hbar \,v_F
%\sqrt{
k_y)^2+m^2 v_F^4]^{1/2} + U_0$ and $E = [(\hbar \,v_F  k_y)^2+m^2
v_F^4]^{1/2}$. At smaller wavevectors, %the
tunneling across the barriers introduces a cut-off in the spectrum
for $E < -[(\hbar \,v_F k_y)^2+m^2 v_F^4]^{1/2} + U_0$. For
confinedstates, the spinor components decay exponentially in the
region
$\xi < - 1/2$. Then the A component %in this region can %then
can be written
as $\phi_A(\xi)=A_1\,e^{\alpha \xi}$. %By a direct s
Substituting $\phi_A(\xi)$
in Eq. (4) we find $\phi_B(\xi)=if_- \,A_1\,e^{\alpha \xi}$, with %here
$f_-=(\beta-\alpha)/(\epsilon-u_0+\Delta)$ and the decay constant
$\alpha$ given by $\alpha = [\beta^2 - (\epsilon - u_0)^2 +
\Delta^2]^{1/2}$, where $u_0 = U_0 L/\hbar v_F$.

The solutions %for
$\phi_A$ and $\phi_B$  for $ |\xi| \leq 1/2$
%in the region $-1/2 \leq\xi \leq 1/2$
are of the type
\begin{eqnarray} %{equation}
\phi_A(\xi)&=&C_2\,\cos(\kappa \xi)+D_2\,\sin(\kappa \xi),\\
%\end{equation}
%\begin{eqnarray}
\phi_B(\xi)&=& %\frac{i}{
[i/(\epsilon+\Delta)]\Bigl[C_2[\beta\,\cos(\kappa \xi)+\kappa
\,\sin(\kappa \xi)]+\cr
&&%\frac{i}{(\epsilon+\Delta)}
\hspace*{1.7cm}D_2\,[\beta\,\sin(\kappa \xi)-\kappa \cos(\kappa
\xi)]\Bigr],
\end{eqnarray}
with $\kappa^2 = \epsilon^2 - \beta^2-\Delta^2$. For $\xi > 1/2$
the solutions are similar to those for %obtained for region
$\xi <-1/2$ but with a negative exponent:
$\phi_A(\xi)=A_3\,e^{-\alpha \xi}, \quad \phi_B(\xi)=if_+
\,A_3\,e^{-\alpha \xi}$, where
$f_+=(\beta+\alpha)/(\epsilon-u_0+\Delta)$. It should be stressed
that, in contrast with the non-relativistic case, the spinor
components are neither even nor odd functions, despite the
symmetry of the potential. This symmetry, however, is reflected in
the probability density $ \rho = \Psi^\dagger \Psi =
\phi_A(\xi)^\dagger \phi_A(\xi) + \phi_B(\xi)^\dagger \phi_B(\xi)$
\cite{Vinc} , which is an even function. Moreover, for a
%in the particular case of the
step potential the derivatives of the spinor components are not
continuous because $u'$ in Eq. (5) is a delta function. This can
be demonstrated by considering the continuity of the $y$ component
of the probability current, $j_y =
v_f\Psi^\dagger \sigma_y \Psi$, across the potential interface: % and
using Eqs. (3) and (4) we obtain ($u_{+}=u_0/(\epsilon +\Delta)$)
\begin{equation}
\hspace*{-0.2cm}{\phi_A'}_\leftarrow(1/2)=%\Bigl
(1-u_{+}%\frac{u_0}{\epsilon +\Delta}
){\phi_A'}_\rightarrow(1/2) + u_{+}\beta \phi_A(1/2),
\end{equation}
where the arrows indicate the limiting values
 from the left and right of the interface. %It is interesting to
Notice in Eq. (8) that, even for large values of $\Delta$, a
continuous derivative of $\phi_A$ may be assumed only for $u_0
\beta<< \Delta$.

Requiring the continuity of $\phi_A$ and $\phi_B$ at $\xi=-1/2$
and $\xi=1/2$ we %are able to
obtain the following transcendental equation for the energy
eigenvalues %($\beta=k_{y}L, \delta=\tan(\kappa L/2)$)
\begin{equation}
\hspace*{-0.1cm}S_-(\epsilon,\beta, +1)\ S_+(\epsilon,\beta, +1)\
+\ S_-(\epsilon,\beta, -1)\ S_+(\epsilon,\beta, -1)=0,
\end{equation}
where $S_\pm( \epsilon,\beta, s)=\beta- f_{\pm}(\epsilon
+\Delta)-s\kappa \delta^{\mp s}$ and $\delta=\tan(\kappa /2)$. The
non-relativistic limit can be obtained using $\epsilon =
\epsilon_c + \Delta$, where $\epsilon_c$ corresponds to the
classical energy and considering the limit $\Delta
>> \epsilon_c$, to give
\begin{equation}
f_\pm(\epsilon + \Delta) \approx (\beta \pm \alpha)( 1+ \Gamma ),
\end{equation}
where $ \alpha \approx [\beta^2+2\Delta(u_0-\epsilon_c)]^{1/2}$,
$\kappa \approx %\sqrt{
(2\Delta \epsilon_c - \beta^ 2)^{1/2} $, and $\Gamma \equiv
u_0/2\Delta$. Equation (9) then becomes ($\bar{\kappa}=\kappa/2$)
\begin{equation} %{eqnarray}
%\Bigl
%\hspace*{-0.1cm}
[ \alpha (1+\Gamma) - \kappa \tan \bar{\kappa}] \ %(\kappa
%/2)] %\,\Bigr]\Bigl
[ \alpha (1+\Gamma) + \kappa \cot \bar{\kappa}] %(\kappa
%/2)]%\,\Bigr]
%\cr &&\cr &&
-2(\beta \Gamma)^2=0.
\end{equation} %{eqnarray}
For $\Gamma << 1$ and $\beta u_0 << \Delta$ we recover the
familiar transcendental equation for
a non-relativistic %electrons in a
QW. % quantum well
% but only if we further assume
%that $\beta u_0 << \Delta$ is also satisfied.
In this limit a non-zero value for $\beta$ is equivalent to a
simple shift of the energy scale $\epsilon' \rightarrow \epsilon_c
- \beta^2/2\Delta$
and the %energy
spectrum of the confined states %corresponds to
becomes a set of nested parabolas. On the other hand, Eq. (11)
shows that, even for massive particles, the QW spectrum does
depend on the $y$ component of the momentum, in contrast with the
non-relativistic results. Thus, a significant modification of the
parabolic spectrum occurs as $\beta$ increases.

Equation (9) was solved numerically.  %and t
The results are shown in Fig. 2 for $U_0 = 50$ meV, $L = 200$
nm, and $\Delta = 0$. The dashed lines delimit the continuum
region, which corresponds to
%%%%%%
free-electrons
%%%%%%
($E \geq \hbar k_y +
U_0$) with energies greater than the barrier height, and
free-holes ($E \leq -\hbar v_F k_y + U_0$) that propagate in the
system by means of the Klein tunneling mechanism. The cut-off at
low wavevectors thus arises due to the conversion of confined
electrons to free holes. For large values of $k_y$ the dispersion
branches are given approximately by
\begin{equation}
E = \hbar v_F  [(n\pi/L)^2+k_y^2]^{1/2},
\end{equation}
where $n$ is an integer. For any given $k_y$, the accuracy of this
approximation improves as $L$ increases.  The lower inset shows
(a) $\phi_A$
%%%%%%
(solid curve) and
$i\phi_B$ (dashed curve)
%%%%%%
for the confined state, with $k_y=0.03$ nm$^{-1}$, shown
by the solid triangle and (b) the corresponding probability
density in arbitrary units. The plot clearly indicates a
discontinuity in the derivative of the spinor component functions
at the barrier interfaces. The vertical dotted lines indicate the
walls of the well.
%We verified these results with a numerical
%integration of Eqs. (3)-(4).
The upper inset shows the effect of the mass, with $m v_F^2 = 10$
meV. The dashed lines again represent the limits of the
free-particle continua. In this case, confined states are allowed,
for $k_y = 0$, in the range $u_0 - \Delta < \epsilon < u_0 +
\Delta$. This energy range broadens as $k_y$ increases and remains
constant for $k_y > (u_0^2/4-\Delta^2)^{1/2}$. At lower energies,
there is again a cut-off, due to the Klein tunnelling at the
barriers, which disappears for $2\Delta > u_0$.
\begin{figure}
\vspace{-1.42cm}
\centering{\resizebox*{!}{6.75cm}{\includegraphics{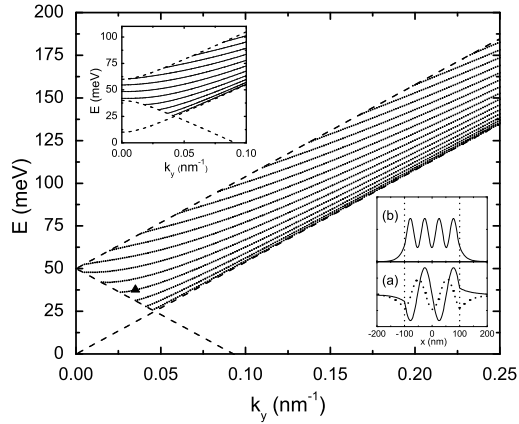}}}
%\centering{\resizebox*{!}{7.25cm}{\includegraphics{newFig2.eps}}}
\vspace{-0.5cm} \caption{Spectrum of confined states in a graphene
{\it square} QW vs $k_{y}$ for $U_0 = 50$ meV, $L = 200$ nm, and
$m v_F^2 = 0$. The lower inset shows (a) $\phi_A$ and $i\phi_B$
for the state shown by the solid triangle and (b) the related
probability density. The upper inset shows the effect of a
non-zero mass, for $m\, v_F^2 = 10$ meV.}
 \label{fig:f2}
\end{figure}

Next, we consider  a QW
with a %continuous
{\it parabolic} potential profile $U(x)=U_0( 2x/L)^2$ for $ |x|
\leq L/2$ and $U(x)=U_0$ for $ |x| > L/2$. Figure 3 shows the
spectrum of confined states obtained from a numerical solution of
Eqs. (3) and (4) for $U_0 = 50$ meV and $L = 200$ nm. The results
are qualitatively similar to those of the previous case, but now
with the eigenvalues being approximately equally spaced for large
wavevectors.
\begin{figure}
\vspace{-1.4 cm}
\centering{\resizebox*{!}{6.75cm}{\includegraphics{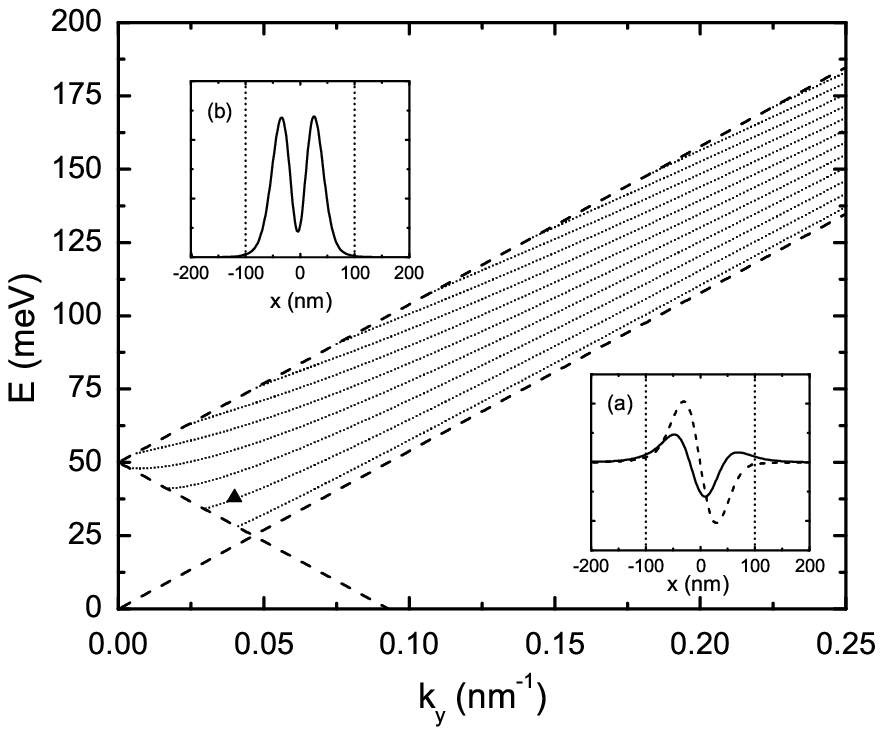}}}
%\centering{\resizebox*{!}{7.25cm}{\includegraphics{newFig3.eps}}}
\vspace{-0.5cm} \caption{ As in Fig. 2 but for a {\it parabolic}
QW.}
\end{figure}

An essential difference with non-relativistic electrons, evident
in all cases, is the appearance of new confined states at the
edges of the continua, where the quantized electron branches
intercept the free-particle regions. Thus, by an adiabatic
increase in $k_y$ one can transform a free-electron or a free-hole
state into a bound electron state. This occurs because the
presence of the barriers allows a mixing of electron and hole
states with the same energy and $y$ component of momentum. As a
result there is constructive interference between confined states
and unbound electron or hole states that are resonantly
transmitted across the QW. We demonstrate this by calculating the
transmission coefficient of electrons incident on a square well.
Consider the propagating solutions $\psi_A(x,y)=\phi_A(x)e^{ik_y
y}$, with
\begin{equation}
\phi_A(x)=\left\{
\begin{array}{ccc}
e^{i\alpha \xi} + B_1 e^{-i\alpha \xi} && \xi < -1/2, \\
&&\\
A_2e^{i\kappa \xi}+B_2e^{-i\kappa \xi} && -1/2 \leq \xi \leq 1/2,\\
&&\\
A_3e^{i\alpha \xi} && \xi > 1/2,
\end{array}
\right.
\end{equation}
where  $\alpha =  [(\epsilon - u_0)^2-\beta^2 - \Delta^2]^{1/2}$;
 the  solutions for $\phi_B$ are obtained as in the
previous calculation.
\begin{figure}
\vspace{-1.2cm} \hspace*{-1cm}
\centering{\resizebox*{!}{6.5cm}{\includegraphics{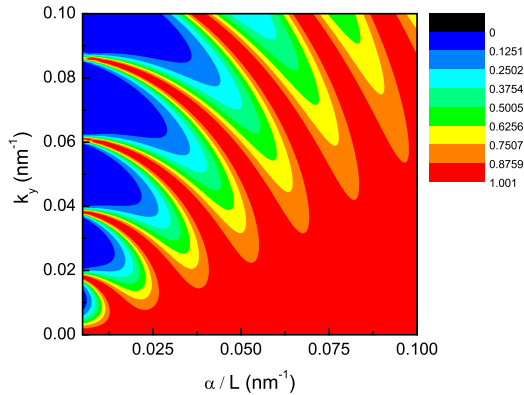}}}
%\centering{\resizebox*{!}{7.5cm}{\includegraphics{newFig4.eps}}}
\vspace*{-0.5cm} \caption{Contour plot of the transmission
coefficient of electrons incident on  a graphene  square well,
with energy $E
> U_0$, as a function of $\alpha$, for $U_0 = 50$ meV, $L =
200$ nm and $m = 0$. }
%\label{fig:wn}
\end{figure}
Then, the transmission coefficient is obtained as $T = |A_3|^2$,
where
\begin{equation}
A_3 = \frac{(g_+-g_-)(f_+-f_-)e^{-i\alpha
}}{(g_+-f_+)(g_--f_-)e^{2i\kappa}-(g_+-f_-)(g_--f_+)},
\end{equation}
$g_\pm = (\beta\pm i\alpha)/(\epsilon + \Delta)$ and  $f_\pm =
(\beta \pm i\kappa)/(\epsilon -u_0+ \Delta)$. A ($k_y, \alpha /L$)
contour plot of the transmission $T$ is shown in Fig. 4 for $U_0
= 50$ meV and $L = 200$ nm. As seen, $T$ depends on the direction
of propagation and displays an oscillatory behavior. As $\alpha
\rightarrow 0$, $T$ reaches a series of maxima for values of
$\beta$ that coincide with the wavevectors for which mixing
occurs, cf. Fig. 2.
%%%%%%%
Notice that for a significant range of
incident angles $T$ is always equal to
$1$. This includes the case
of nearly normal incidence, $k_{y}\approx 0$, and is in sharp
contrast with the non-relativistic case in which $T$ exhibits
periodic maxima equal to $1$ as a function of $k_{x}$. A similar
%%%%%%
direction-dependent transmission  %is to that
through %a potential
graphene barriers was reported recently \cite{kat}. A
direction-dependent transmission is also possible for
non-relativistic electrons tunneling through magnetic barriers
\cite{mat}.

The $y$ components of the momentum for which mixing is allowed
correspond to confined states for which the asymptotic limit
$\alpha \rightarrow 0$ applies. This yields the condition
$\sin{(\kappa )}= 0$ or $\kappa = n \pi$, where $n$ is an integer.
Using the definition of $\kappa$ and $\alpha$
%%%%%%that
gives
\begin{equation}
\beta = % \sqrt{
\Bigl[\Bigl(
\frac{n^2\pi^2}{2u_0}-\frac{u_0}{2}\Bigr)^2-\Delta^2\Bigr]^{1/2}.
\end{equation}
Since $\beta^2 > 0$, the values of $n$ can be obtained from the
condition $\pm (n^2 \pi^2/2u_0- u_0/2)\geq \Delta$, where the $+$
($-$) sign is associated with the upper (lower) continuum edges.
From this condition we find that for $U_0 < 2 mv_F^2$ there is no
mixing at lower energies, although it persists at the upper
continuum edge and the minimum value of $\beta$ for the mixing
increases with $\Delta$.

A complementary way to see the direction dependence of the
transmission $T$ is shown in Fig. 5(a), with $T$ plotted versus
the angle of incidence $\theta = \arctan{(k_y/\alpha)}$, for
different electron energies as indicated. The QW parameters are
$L=200$ and $U_0 =50$ meV.  Notice that for $\theta\approx 0$, we
have $T\approx 1$
%the transmission is approximately 1
in agreement with the $k_{y}\approx 0$ part of Fig. 4. In Fig.
5(b) we plot  $T$ versus the energy $E$ for $\theta = \pi/3$. As
seen, $T$ oscillates with the energy due to the resonance effect
caused by the confined states (as in the Ramsauer-Townsend
effect). The energies for the maxima of the transmission can  be
obtained from Eq. (15) as $\epsilon = (n\pi)^2/2u_0 + u_0/2$.

In summary,  we showed that it is possible to  confine massless
charge carriers by means of electrostatic potentials, due to the
wavevector-dependent suppression of the electron-hole conversion
at the potential barriers. We thus obtained the quantized spectrum
of confined electron states in graphene quantum wells (QW) as a
function of the $y$ component of the wavevector. The results show
a remarkable dependence of the eigenvalues on the momentum with a
cut-off at low wavevectors. The relativistic correction to the
classical QW spectrum leads to a wavevector dependence of the
number of confined states due to the electron-hole conversion at
the continuum edges. Accordingly, such QWs must be treated as
inherently 2D systems.  This is further demonstrated by the
directional dependence of the transmission shown in Figs.  4 and 5.
Studying the resonance transmission of electrons across a QW with
energies above the height of the confining walls, $E > U_0$, can
probe the discrete levels which can be populated by tuning the
Fermi energy of the system with the electric-field effect
\cite{Novo3}.
\begin{figure}
\vspace{-0.5cm}
\hspace*{-0.4cm}\centering{\resizebox*{!}{6.5cm}{\includegraphics{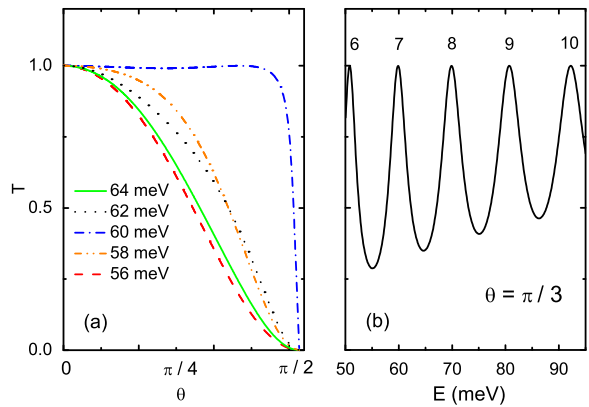}}}
\vspace{-0.7cm} \caption{ (a) Transmission $T$ vs angle of
incidence $\theta$ for different energies as indicated. (b) $T$ vs
energy E for $\theta = \pi/3$. The maxima are marked by the
numbers of the confined states.}
\end{figure}

This work was supported by the Brazilian Council for Research
(CNPq), the Flemish Science Foundation (FWO-Vl), the Belgian
Science Policy (IUAP) and the Candian NSERC Grant No. OGP0121756.


\begin{thebibliography}{9}

\bibitem{Novo3}
K. S. Novoselov {\it et al.}, Science, {\bf 306}, 666 (2004).

\bibitem{Novo2}
K. S. Novoselov {\it et al.}, PNAS {\bf 102}, 10451 (2005).

\bibitem{Zhang1}
Y. Zhang {\it et al.}, Appl. Phys. Lett. {\bf 86}, 073104 (2005).

\bibitem{Zheng}
Y. Zheng and T. Ando, Phys. Rev. B {\bf 65}, 245420 (2002).

\bibitem{Sharapov}
V. P. Gusynin and S. G. Sharapov, Phys. Rev. Lett. {\bf 95},
146801 (2005).

\bibitem{Novoselov}
K. S. Novoselov {\it et al.}, Nature {\bf 438}, 197 (2005).

\bibitem{Zhang}
Y. Zhang {\it et al.}, Nature {\bf 438}, 201 (2005).

\bibitem{Zhang2}
Y. Zhang, J. P. Small, M. E. S. Amori and P. Kim, Phys. Rev. Lett.
{\bf 94}, 176803 (2005).

\bibitem{High}
J. Reinhardt and W. Greiner, Rep. Prog. Phys. {\bf 40}, 219
(1977); V. Petrillo and Davide Janner, Phys. Rev. A {\bf 67},
012110 (2003)

%\bibitem{Petrillo}
%V. Petrillo and Davide Janner, Phys. Rev. A {\bf 67}, 012110
%(2003).

\bibitem{Wallace}
P. R. Wallace, Phys. Rev. {\bf 71}, 622 (1947); M. Wilson, {\it
Physics Today}, January 2006, p. 21.

\bibitem{Semenoff}
G. W. Semenoff, Phys. Rev. Lett. {\bf 53}, 2449 (1984).

\bibitem{Kopele}
I. A. Luk'yanchuk and Y. Kopelevich, Phys. Rev. Lett. {\bf 93},
166402 (2004).

\bibitem{Kane}
C. L. Kane and E. J. Mele, Phys. Rev. Lett. {\bf 95}, 226801
(2005).

\bibitem{Vinc}
D. P. DiVincenzo and E. J. Mele, Phys. Rev. B {\bf 29}, 1685
(1984).

\bibitem{Klein}
O. Klein, Z. Phys. {\bf 53}, 157 (1929).

\bibitem{Calogero}
N. Dombey and A. Calogeracos, Phys. Rep. {\bf 315}, 41 (1999).

\bibitem{Peres} N. M. R. Peres {\it et al.}, %A. H. Castro Neto and F. Guinea,
(cont-mat/0604323); J. Tworzyd{\l}o {\it et al.},
(cond-mat/0603315).

\bibitem{kat} M. I. Kattsnelson {\it et al.}, (cond-mat/0604323); V. V. Cheianov and V. FalÕko (cond-mat/0603624).

\bibitem{mat}  A.  Matulis, F. M. Peeters, and P. Vasilopoulos, Phys. Rev. Lett. {\bf 72}, 1518 (1994).
\end{thebibliography}
\end{document}